\documentclass[conference]{IEEEtran}
\IEEEoverridecommandlockouts
\usepackage{cite}
\usepackage{amsmath,amssymb,amsfonts}
\usepackage{algorithmic}
\usepackage{graphicx}
\usepackage{textcomp}
\usepackage{xcolor}
\usepackage{url}
\renewcommand{\IEEEkeywordsname}{Keywords}

\begin{document}

\title{Can Bandits Learn to Share a Channel? Distributed Multi-Armed Bandits for Slotted ALOHA\\
\thanks{This research was funded by the National Science Centre, Poland (2023/05/Y/ST7/00004). We gratefully acknowledge the support of the National Research Institute, grant number POIR.04.02.00-00-D008/20-01 on ``National Laboratory for Advanced 5G Research'' (acronym PL-5G) for providing computing facilities.
\textit{(Corresponding author: Kamil Szczech.)}}
}

\author{
\IEEEauthorblockN{Kamil Szczech}
\IEEEauthorblockA{\textit{AGH University of Krakow}\\
Krakow, Poland\\
kamil.szczech@agh.edu.pl}
\and
\IEEEauthorblockN{Maksymilian Wojnar}
\IEEEauthorblockA{\textit{AGH University of Krakow}\\
Krakow, Poland\\
maksymilian.wojnar@agh.edu.pl}
\and
\IEEEauthorblockN{Krzysztof Rusek}
\IEEEauthorblockA{\textit{AGH University of Krakow}\\
Krakow, Poland\\
krzysztof.rusek@agh.edu.pl}
\and[\linebreak\mbox{}\hfill]
\IEEEauthorblockN{Katarzyna Kosek-Szott}
\IEEEauthorblockA{\textit{AGH University of Krakow}\\
Krakow, Poland\\
katarzyna.kosek-szott@agh.edu.pl}
\and
\IEEEauthorblockN{Szymon Szott}
\IEEEauthorblockA{\textit{AGH University of Krakow}\\
Krakow, Poland\\
szymon.szott@agh.edu.pl}
}

\maketitle

\begin{abstract}
Slotted ALOHA networks constitute a playground for research on machine learning (ML)-based channel access schemes. We consider ML agents deployed in such networks and operating under severe constraints: fully distributed (independent operation), fully online (no pre-training), aperiodic (no fixed schedule), and with no explicit coordination (no signaling). We ask whether a multi-armed bandit (MAB), a minimal reinforcement learning agent, can learn the optimal policy for a capture-free slotted ALOHA system. We evaluate several MAB algorithms on aggregate network throughput and per-node fairness. Our best-performing agent is based on a Softmax bandit, which we call SALSA (Softmax ALOHA Learning for Shared Access). SALSA learns a per-node transmission policy close to optimal, with throughput near the theoretical maximum for fair random access. We achieve performance comparable to a competing state-of-the-art solution, but without requiring a priori knowledge of the network size.
\end{abstract}


\section{Introduction}

Wireless channel access networks have evolved considerably since the original ALOHA protocol \cite{abramson1977throughput}.
Despite the undeniable success of IEEE 802.11 (Wi-Fi) and cellular networks, slotted ALOHA access remains in use in the Internet of Things (IoT) \cite{seo2025aloha} and satellite communications \cite{fidele2026analytical}. 
With the emergence of machine learning (ML), slotted ALOHA networks have become an attractive testbed for evaluating ML-based channel access schemes \cite{zhao2025throughput}.
While the operating limits of capture-free slotted ALOHA are known analytically,\footnote{Under full buffer and perfect node fairness, slotted ALOHA approaches maximum throughput at $\frac{1}{e}$ as the network size increases \cite{abramson1977throughput}.} we ask whether ML agents can rediscover efficient behavior under severe constraints: in a fully distributed environment without explicit communication and without relying on scheduled (cyclic) operation. In particular, we evaluate minimal reinforcement learning (RL) agents: multi-armed bandits (MABs), which operate fully online (without pre-training).

So far, MABs have been mostly used in research on wireless networks to optimize channel access parameters or to perform resource allocation \cite{10772999, casasnovas2025learning, raza2024multi, peng2024multi}. Most relevant to our work, Zhao et al. \cite{zhao2025throughput} design MAB-based throughput-optimal access with local rewards (MTOA-L) for both capture-free and capture-based systems. The latter (capture-based MTOA-L) achieves high throughput but suffers from short-term unfairness. In contrast, capture-free MTOA-L requires \textit{a priori} knowledge of the network size. Our goal is to maintain high short-term fairness without knowing the network size.

Our contributions are the following. First, we show that under our operating assumptions (no mobility, a fixed set of nodes, and full buffers), distributed channel access reduces to a multi-agent, non-contextual bandit problem (Section~\ref{sec_problem_formulation}). Second, we implement several MAB agents along with state-of-the-art baselines (Section~\ref{sec_baselines}) in an open-source channel access simulator\footnote{\url{https://github.com/mldr-devs/ltc/tree/mab}} (Section~\ref{sec_simulator}).
Third, we evaluate our preselected and hyperparameter-tuned group of MAB algorithms (Section~\ref{sec_mab_evaluation}) and determine that Softmax-based agents perform the best.
We call this approach SALSA (Softmax ALOHA Learning for Shared Access).
Fourth, we evaluate SALSA against selected baselines (Section~\ref{sec_comparison}) and study the impact of PHY layer errors on convergence and network performance (Section~\ref{sec_channel_errors}).
We find that SALSA operates close to the theoretically optimal throughput of fair channel access, i.e., $\frac{1}{e}$, while ensuring high per-node fairness, without requiring \textit{a priori} knowledge of the network size.
Finally, in Section~\ref{sec_conclusions}, we conclude our study and outline future work.

\section{System Model}

\subsection{Problem Formulation}
\label{sec_problem_formulation}

The network consists of $N$ independent agents that share a common wireless channel and operate in synchronous time slots indexed by $t=0,1,\dots,\allowbreak T-1$. Each agent is fully autonomous: it runs its own learning and decision making, the network has no central controller, and no control messages. We further assume full-buffer traffic, so that every node always has data to send, no mobility, and a fixed number of nodes throughout the experiment. Thus, the problem involves no state transitions and takes the form of a bandit problem. The agents are trained separately and optimize the same reward function, which makes the interaction a non-cooperative game in which one agent's payoff is coupled to the choices of the others. Therefore, from the perspective of an individual agent, the environment is not stationary, since the channel state is shaped by the evolving policies of other agents. We investigate whether the agents find an equilibrium point, maximizing the aggregate network throughput while achieving high fairness.

We formulate the problem as a multi-agent, non-contextual bandit problem with $N$ agents. No agent observes the actions or rewards of the others, and the goal of each agent is to maximize its own cumulative reward over the time horizon of $T=30\,000$ slots. At every slot $t$, each agent $i$ selects one action $a_i^t\in\mathcal{A}=\{\mathrm{TX},\ \mathrm{ID}\}$, corresponding to transmit or idle. When the agent selects $\mathrm{TX}$, let $\tau_i^t\in\{-1,1\}$ denote the transmission outcome\footnote{Following other state-of-the-art papers, we assume that acknowledgments are sent back immediately and that their transmission is part of the slot duration.}, where $\tau_i^t=1$ marks a successful transmission and $\tau_i^t=-1$ a collision. The reward is then
\begin{equation}
R_i^t = \begin{cases}
1,   & a_i^t=\mathrm{TX} \land \tau_i^t=1, \\
0,   & a_i^t=\mathrm{ID}, \\
-1,   & a_i^t=\mathrm{TX} \land \tau_i^t=-1.
\end{cases}
\label{eq:reward_tx}
\end{equation}
Thus, the agent is rewarded for a successful transmission but penalized for colliding.

\subsection{Network Simulator}
\label{sec_simulator}
We implement the system in a custom Python-based network simulator, using  JAX to accelerate execution. Our simulator follows a slotted (Monte Carlo) model with abstract, unitless time intervals. Within each slot, exactly one of three channel states occurs: a successful packet transmission, a collision, or an idle channel. In addition, random channel errors can be configured following a Bernoulli process with a specified error probability. Such errors are indistinguishable from a collision. We assume all nodes lie within each other's transmission range (no hidden nodes) so no channel attenuation is required.

Our main metric is throughput defined as the ratio of the number of slots with successful transmissions to all slots. We use aggregate throughput (of all nodes) to determine network performance and per-node throughput $x_i$ as input for Jain's fairness index, $J = (\sum_{i} x_i)^2 / (N \sum_{i} x_i^2) \in [1/N, 1]$, to assess throughput distribution, where $J=1$ denotes a perfectly equal share.

\subsection{MAB Agents}
\label{sec_mab_agents}
We apply well-known MAB algorithms to the problem: Thompson sampling (TS), EXP3~\cite{auer2002nonstochastic}, and Softmax (i.e., the gradient bandit)~\cite{sutton2018reinforcement}. Two variants of TS are used: continuous and discrete. The former models the reward with a Gaussian likelihood and a normal-inverse-gamma conjugate prior~\cite{murphy2007conjugate}. The latter uses a Dirichlet prior, which is the conjugate prior of the categorical reward distribution (since the reward takes only one of three values: $-1$, $0$, and $1$)~\cite{tu2014dirichlet}. Softmax is implemented with a temperature parameter that controls the exploration and exploitation trade-off, together with an exponential recency-weighted average of the rewards. The continuous TS additionally discounts its posterior parameters over time, preventing it from collapsing into an $\arg\max$ policy.

Algorithm hyperparameters are tuned to maximize both throughput and fairness and are uniform across all agents in the network. We perform the tuning with Optuna~\cite{akiba2019optuna} on networks of $5$ and $20$ agents, and the resulting values are reused in all experiments throughout the paper (Table~\ref{tab:parameters}).

\begin{table}[t]
\caption{Hyperparameters of the MAB agents.}
\begin{center}
\begin{tabular}{|c|c|c|l|}
\hline
\textbf{Algorithm} & \textbf{Parameter} & \textbf{Value} & \textbf{Description} \\
\hline
EXP3 & $\gamma$ & $0.001$ & Exploration factor \\
\hline
Softmax & $\eta$ & $0.02$ & Step size \\
 & $\alpha$ & $0.93$ & Recency weight \\
 & $\tau$ & $9.9$ & Temperature \\
\hline
Continuous TS & $\alpha$ & $0.75$ & Inverse-gamma shape \\
 & $\beta$ & $0.06$ & Inverse-gamma scale \\
 & $\lambda$ & $0.004$ & Pseudo-observation count \\
 & $\mu$ & $-0.2$ & Prior mean \\
 & $\gamma$ & $0.96$ & Discount factor \\
\hline
Discrete TS & $\alpha$ & $94.0$ & Dirichlet concentration \\
\hline
\end{tabular}
\label{tab:parameters}
\end{center}
\end{table}

\subsection{Baselines}
\label{sec_baselines}
We evaluate our approach against three baselines:

\begin{itemize}
    \item \textbf{Exponential Backoff ALOHA (EB-ALOHA)}~\cite{yu2019deep}: A standard slotted ALOHA baseline with a contention window range of $[4, 16]$. 
    \item \textbf{ALOHA-Q}~\cite{chu2012aloha}: A Q-learning MAC protocol where autonomous agents learn to occupy reserved slots within synchronized transmission cycles. As the active node count $N$ is unknown \textit{a priori}, we use a fixed cycle length of $C = 20$ slots across all network sizes.
    \item \textbf{MTOA-L}~\cite{zhao2025throughput}: The distributed variant of MTOA, where nodes select between a transmission action and $L$ null actions. We evaluate both its capture-free and capture-based variants. For capture-free operation, optimal throughput requires $L = N - 1$. Since $N$ is unknown \textit{a priori}, we set $L = 19$ (matching $N = 20$) as a fixed reference.
\end{itemize}
The first baseline is a heuristic and approximates channel access in IEEE 802.11 networks. The remaining baselines are state-of-the-art competing solutions.

\section{Performance Evaluation}

We evaluate the performance of the selected MAB agents, present the policies they learn, compare our best-performing agent against the baselines, and assess its robustness under a noisy channel. We present the means of five independent runs as well as 95\% confidence intervals.

\subsection{MAB Evaluation}
\label{sec_mab_evaluation}

\begin{figure}
    \centering
    \includegraphics[width=0.8\linewidth]{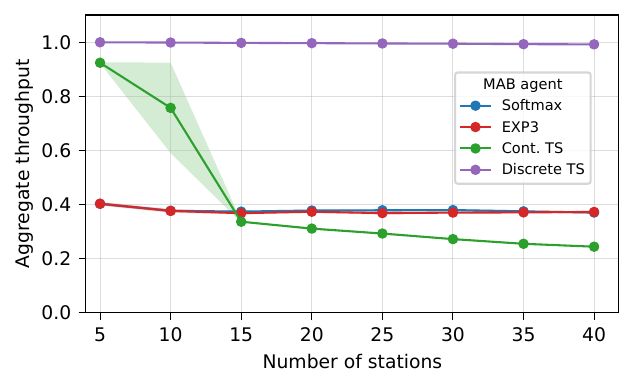}
    \caption{Aggregate throughput of MAB agents. Results averaged over the last 10\% of slots.}
    \label{fig:thr_mab}
\end{figure}

Fig.~\ref{fig:thr_mab} shows the aggregate throughput achieved by each
MAB agent as a function of the number of stations $N$ in the network.
For large $N$, Softmax and EXP3 remain close to $1/e \approx 0.368$, i.e., the maximum theoretical throughput of fair random channel access \cite{abramson1977throughput}. 
Continuous TS achieves high throughput for a small number of stations (up to $\sim$0.92 at $N=5$), but this drops sharply between $N=5$ and $N=15$ with substantial variance across seeds, and continues to decline gradually thereafter, falling below Softmax and EXP3. Discrete TS maintains throughput close to~1 regardless of $N$.

\begin{figure}
    \centering
    \includegraphics[width=0.8\linewidth]{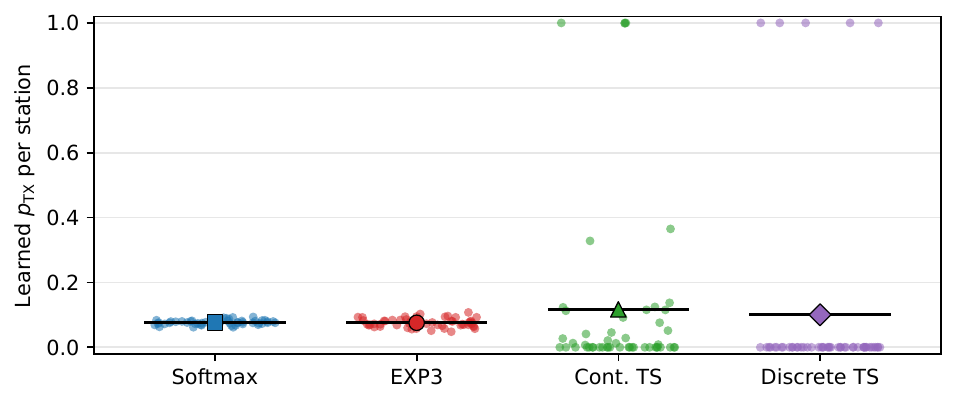}
    \caption{Learned per-node transmission policy $p_\mathrm{TX}$ for $N=10$ nodes across all seeds. Each point is one node in one run; the highlighted marker denotes the mean value for each agent type. The optimal policy is $p_\mathrm{TX}=1/N$.}
    \label{fig:policy_n10}
\end{figure}

Since aggregate throughput alone does not reveal how evenly this throughput is shared among stations, we examine the per-station policies learned by each agent. Fig.~\ref{fig:policy_n10} shows that Softmax and EXP3 agents converge to nearly identical transmission probabilities $p_\mathrm{TX}$ across all stations, so the resulting throughput is distributed almost equally. However, the variance of the Softmax policy is slightly lower, implying a higher Jain's fairness index than EXP3.
Conversely, agents using continuous and discrete TS learn a highly bimodal policy; most stations converge to $p_\mathrm{TX} \approx 0$, while a few stations (or even just one for discrete TS) reach $p_\mathrm{TX} \approx 1$. 
Thus, a small subset of stations captures almost all the channel access opportunities, while the remaining stations almost never transmit, which translates into low fairness.

In general, among the MAB agents evaluated, Softmax achieves the best trade-off between throughput and fairness. Henceforth, we refer to our solution with Softmax agents as SALSA and evaluate it against the literature baselines.

\subsection{Comparing SALSA with Baselines}
\label{sec_comparison}

\begin{figure}
    \centering
    \includegraphics[width=0.85\linewidth]{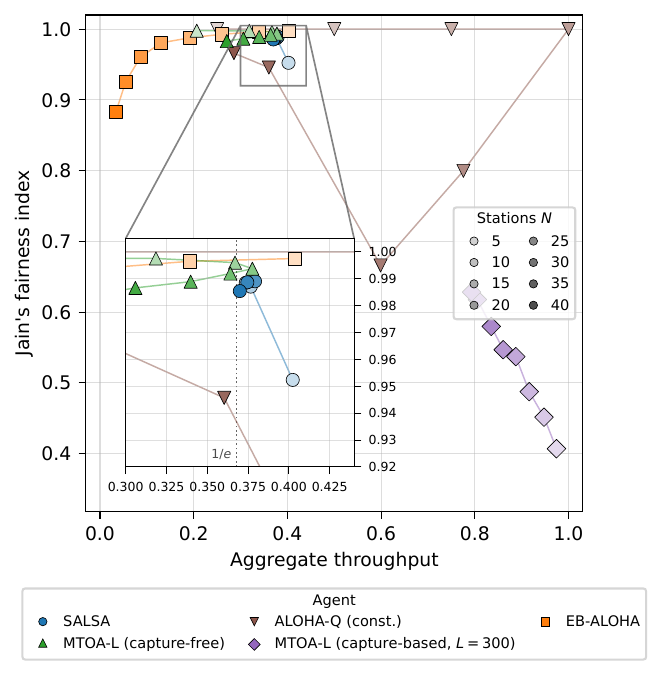}
    \caption{Pareto front of aggregate throughput vs. Jain's fairness index for the proposed SALSA and baseline channel-access algorithms as a function of the number of nodes. Partially overlapping points around $1/e$ have been enlarged to improve readability. Results averaged over the last 10\% of slots.}
    \label{fig:pareto_baselines}
\end{figure}

Fig.~\ref{fig:pareto_baselines} presents the Pareto front, comparing the combined throughput and fairness of SALSA with baseline channel access schemes. The performance of EB-ALOHA and ALOHA-Q is highly dependent on the number of nodes in the network. EB-ALOHA's performance declines as the number of nodes increases, and the algorithm operates close to the optimum only when the number of nodes is 5 or 10, due to the fixed contention window range used by the algorithm. Similarly, ALOHA-Q operates at full efficiency only when $N=20$, matching its fixed cycle length; otherwise throughput drops.

Another technique that requires \textit{a priori} knowledge of the number of nodes is capture-free MTOA-L. With $L = N-1$, it achieves performance similar to our proposed approach \cite{zhao2025throughput}. In our tests, we use a fixed $L=19$ to illustrate the impact of this setting on MTOA-L operation under a varying number of nodes. We observe that, for $N=20$, fairness is high and throughput is similar to that of SALSA. As $N$ deviates from this optimal value in either direction, MTOA-L throughput likewise degrades.

MTOA-L also offers a capture-based mode that does not require knowledge of the exact number of nodes, using $L \gg N$ (e.g., $L=300$). In this mode, network throughput is close to~1 regardless of the number of nodes in the network. However, this comes at the cost of a drastic drop in the fairness coefficient, since MTOA-L achieves high aggregate throughput at the expense of short-term fairness. In contrast, SALSA provides throughput close to the theoretical maximum for fair channel access (for large $N$), i.e., near $\frac{1}{e}$, while maintaining high fairness (over 0.93).

\begin{figure}
    \centering
    \includegraphics[width=0.8\linewidth]{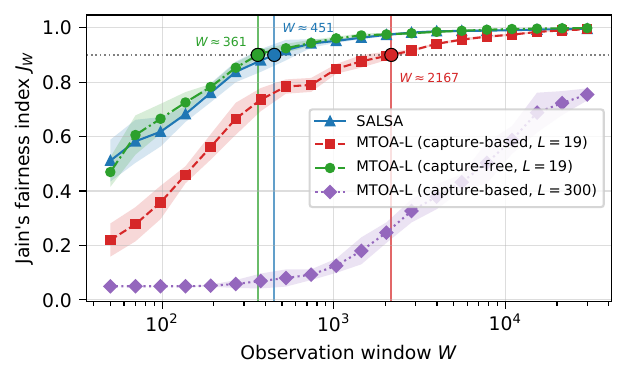}
    \caption{Short-term Jain's fairness index $J_W$ as a function of the observation window $W$ (number of trailing slots) for $N=20$ nodes and no channel errors, for SALSA and three MTOA-L variants. Vertical lines
    mark the window size at which each curve reaches a fairness of 0.9 (horizontal dotted line).}
    \label{fig:fairness_vs_samples}
\end{figure}

A further study of fairness is given in Fig.~\ref{fig:fairness_vs_samples}, which shows Jain's fairness index in the short-term as a function of the observation window $W$ for the network composed of $N=20$ nodes. SALSA and capture-free MTOA-L (with $L=N-1$) reach a fairness of 0.9 the fastest, after $W\approx450$ and $W\approx350$ slots, respectively. Capture-based MTOA-L with $L=N-1$ reaches the same fairness level much slower ($W\approx2200$), indicating that the capture-based resolution mechanism itself decreases short-term fairness even when $L$ is set to its optimal value. Capture-based MTOA-L with $L=300$ does not reach a fairness of 0.9 within the observation range considered, which is consistent with its low fairness observed in Fig.~\ref{fig:pareto_baselines}.

\subsection{Robustness to Channel Errors}
\label{sec_channel_errors}

\begin{figure}
    \centering
    \includegraphics[width=0.8\linewidth]{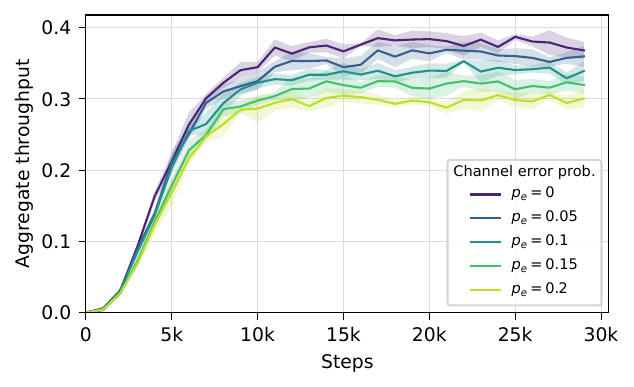}
    \caption{Aggregate network throughput over successive steps for
    SALSA at $N=20$ nodes, under different channel error
    probabilities $p_e$.}
    \label{fig:thr_error}
\end{figure}

Fig.~\ref{fig:thr_error} shows the aggregate network throughput over successive steps for SALSA tested in five independent trials on channels with varying probabilities of channel errors $p_e$. As $p_e$ increases, the maximum achievable network throughput decreases in proportion to $1-p_e$. Furthermore, we do not observe a prolonged convergence time as $p_e$ increases, meaning that SALSA is able to find the optimal solution even in a non-ideal radio
channel, which increases the algorithm's potential for practical application.

\section{Conclusions}
\label{sec_conclusions}

We investigated whether minimal RL agents can learn to equally share a channel. Our answer is affirmative: of the MAB algorithms evaluated, Softmax-based agents (SALSA) converge to near-identical per-node transmission probabilities and operate close to $\frac{1}{e}$, the throughput limit of fair random access, while maintaining a high Jain's fairness index, all without \textit{a priori} knowledge of $N$, without signaling, and without pre-training.
Comparing to the state of the art, SALSA matches capture-free MTOA-L configured with its optimal $L=N-1$ but does not require that configuration, and reaches short-term fairness several times faster than the capture-based variant. SALSA also tolerates imperfect channels, with throughput scaling as $1-p_e$ while having convergence time unaffected.  The main limitation of MABs is that they are stateless so as future work we foresee replacing bandits with deep RL agents, while retaining the same constraints. Relaxing the fixed, saturated, homogeneous population assumption towards dynamic arrivals, non-saturated traffic, and legacy coexistence remains open.

\bibliographystyle{IEEEtran}
\bibliography{bibliography}

\end{document}